\documentclass[a4paper,fleqn]{cas-sc}

\usepackage[authoryear,longnamesfirst]{natbib}
\def\tsc#1{\csdef{#1}{\textsc{\lowercase{#1}}\xspace}}
\tsc{WGM}
\tsc{QE}
\tsc{EP}
\tsc{PMS}
\tsc{BEC}
\tsc{DE}
\usepackage{enumitem}

\begin{document}
\let\WriteBookmarks\relax
\def\floatpagepagefraction{1}
\def\textpagefraction{.001}
\shorttitle{}
\shortauthors{K Sharma et~al.}

\title [mode = title]{Measuring the continuous research impact of a researcher: The $K_z$ index}                      

%

\author[1]{Kiran Sharma} 
\ead{kiran.sharma@bmu.edu.in} 
\author[1]{Ziya Uddin} 
\ead{ziya.uddin@bmu.edu.in}
\address[1]{School of Engineering \& Technology, BML Munjal University,  Gurugram, Haryana-122413, India}
\cortext[cor1, cor2]{Corresponding authors}

\begin{abstract}
The ongoing discussion regarding the utilization of individual research performance for academic hiring, funding allocation, and resource distribution has prompted the need for improved metrics. While traditional measures such as total publications, citations count, and the $h$-index provide a general overview of research impact, they fall short of capturing the continuous contribution of researchers over time. To address this limitation, we propose the implementation of the $K_z$ index, which takes into account both publication impact and age. In this study, we calculated $K_z$ scores for 376 research profiles. $K_z$ reveals that the researchers with the same $h$-index can exhibit different $K_z$ scores, and vice versa. Furthermore, we observed instances where researchers with lower citation counts obtained higher $K_z$ scores, and vice versa. Interestingly, the $K_z$ metric follows a log-normal distribution.  It highlights its potential as a valuable tool for ranking researchers and facilitating informed decision-making processes. By measuring the continuous research impact, we enable fair evaluations, enhance decision-making processes, and provide focused career advancement support and funding opportunities.

\end{abstract}


\begin{highlights}
\item Traditional metrics such as total publications, citations count, and the $h$-index provide an overall measure of research impact but fail to capture the continuous contribution of researchers. Therefore, there is a need for a robust tool to measure the continuous research impact.

\item The proposed $K_z$ index is introduced as a solution, taking into account both the impact and age of publications.

\item Even if two or more researchers have identical total publications, citations count, and $h$-index, it is unlikely that they share the  same $K_z$ scores This characteristic of the $K_z$ index makes it a valuable ranking tool.

\item The $K_z$ index enables the identification of both star contributors and those with lower impact in the realm of research.

\item By measuring the continuous research impact, a more comprehensive assessment can be achieved, leading to fair evaluations towards career progression support and research funding.

\end{highlights}


\maketitle

\section{Introduction}

Research impact is a crucial factor when evaluating the contributions of researchers~\citep{egghe2010hirsch}. It plays a vital role in assessing the quality, significance, and reach of their work, which is instrumental in academic promotions, grant allocations, award selections, and overall career progression. Existing indices like the $h$-index and citation count are commonly used to measure research impact~\citep{bornmann2005does, bornmann2009state}; however, it's important to recognize that citations may not provide a comprehensive representation of impact, especially in fields where citation practices differ or in emerging research domains with limited citation opportunities. Therefore, a more nuanced approach is necessary to capture the full extent of the research impact, considering multiple dimensions beyond traditional metrics.

The $h$-index has been subject to criticism due to its limitations in providing a comprehensive view of scientific impact~\citep{costas2007h}. Initially introduced in 2005 by Hirsch, the $h$-index is calculated based on the number of papers that have received at least $h$ citations from other papers \citep{hirsch2005index}. Since its introduction, the $h$-index has gained significant popularity in academia and has been commonly employed to evaluate the academic success of scientists in various areas, including hiring decisions, promotions, and grant acceptances. Despite efforts by researchers to propose alternative variants of the $h$-index ~\citep{egghe2006theory,jin2007r, zhang2009index,  alonso2010hg, khurana2022impact}, the traditional $h$-index remains widely used as a performance metric in the assessment of scientists because of its simplicity.

To overcome the limitations of $h$-index, Egghe in 2006 proposed $g$-index which is determined by the distribution of citations across their publications.  It is determined by sorting the articles in decreasing order based on the number of citations they have received. The $g$-index is defined as the largest number $g$ for which the top $g$ articles collectively accumulate at least $g^2$ citations~\citep{egghe2006theory}. This means that a researcher with a $g$-index of 10 has published at least 10 articles that collectively have received at least  $(10^2 = 100)$ citations. It's important to note that unlike the $h$-index, the citations contributing to the $g$-index can be generated by only a small number of articles. For example, a researcher with 10 papers, where 5 papers have no citations and the remaining five have 350, 35, 10, 2, and 2 citations respectively, would have a $g$-index of 10 but an $h$-index of 3 (as only three papers have at least three citations each).

Further, after recognizing the limitations of the $h$-index~\citep{ding2020exploring}, researchers have proposed various complementary measures to provide a more comprehensive assessment of research impact such as $R$-index~\citep{jin2007r},  $e$-index~\citep{zhang2009index},  $h'$-index~\citep{zhang2013h}.
In the study by Khurana et al. (2022)~\citep{khurana2022impact}, an enhancement to the $h$-index is proposed to capture the impact of the highly cited paper. They introduced $h_c$ which is based on the weight assigned to the highly cited paper.  $h_c$ has a greater impact on researchers with lower $h$-index values, particularly by highlighting the significance of their highly cited paper. However, the effect of $h_c$ on established researchers with higher $h$-index values was found to be negligible. It is worth noting that the $h_c$ focuses on the first highly cited paper and does not consider the impact of subsequent highly cited papers. This limitation again highlights the need for a more comprehensive measure that takes into account all the important factors contributing to research impact~\citep{martin1996use}.

The another measure named, $L$-sequence, introduced by Liu et al.~\citep{liu2014empirical}, computes the $h$-index sequence for cumulative publications while taking into account the yearly citation performance. In this approach, the $L$ number is calculated based on the $h$-index concept for a specific year. Consequently, the impact of the most highly cited paper in that year may be overlooked, and papers with less than $L$ citations are also not considered. Although the concept captures the yearly citation performance of all papers, it does not effectively capture the continuous impact of each individual paper. Also gathering data for the $L$-sequence can be challenging, as it requires delving into the citation history of each paper for every year.

Quantifying research impact is a multifaceted endeavor~\citep{batista2006possible}. There is no universally accepted metric or methodology for measuring continuous research impact, and different stakeholders may prioritize different indicators, such as publications, citations, patents, or societal impact. Measuring the continuous research impact of a researcher is crucial for granular assessment, differentiation among researchers, funding decisions, identification of emerging talent, etc.  Determining an inclusive and comprehensive approach that captures the diverse dimensions of research impact remains a challenge.

\subsection{Research Objective}
The primary objective of this study is to introduce a reliable metric that can effectively capture the continuous research impact of a researcher. The aim of the proposed metric is to differentiate between two researchers who possess identical research parameters. In order to accomplish the stated objective, a newly introduced measure called the $K_z$-index has been proposed.


\section{$K_z$-index}
\label{method}
The proposed $K_z$ index serves as a tool to measure the research impact of a researcher. It aims to capture the continuous and evolving contributions made by the researcher over time, considering factors such as total publications, citation count, and publication age.
\subsection{Definition of $K_z$-index}

To measure the continuous research impact of a researcher, $K_z$ takes into account two important factors of research:

\begin{enumerate}
\item \textbf{Impact ($k$)}: The impact of a paper is determined by considering two factors: the number of citations ($C$) it has received and its $h$-index.\\

The impact of the paper is calculated by using the following equation;
\begin{equation}
C \leq (h+1)^k ,
\label{eq1}
\end{equation}

where $k \in \mathbb{R}^{+}$.

\item \textbf{Age ($\Delta t$)}: $\Delta t$ represents the publication age in relation to the current year and can be determined through the following computation.
\begin{equation}
\Delta t = C_y - P_{y}
\label{eq2}
\end{equation}

where $C_y$ represents the current year and $P_y$ represents the publication year.

\end{enumerate}

Now,  from Eq.\ref{eq1} and Eq.\ref{eq2}, $K_z$ can be calculated for every researcher $i$ as
\begin{equation}
K_z = \sum_{i=1}^N  \frac{k_i}{\Delta t_i}
\label{eq3}
\end{equation}
 where $N$ is number of publications and $N>0$.

Equation~\ref{eq3} highlights the sig nificance of $K_z$ metric by incorporating essential research indicators, including total citations, year of publication, number of publications, publication age, and $h$-index. This comprehensive approach ensures that all significant aspects of a researcher's work are considered, resulting in a more robust and holistic assessment of their research impact.

\subsection{Advantages of $K_z$}
Measuring the continuous research impact of a researcher is crucial for several reasons:
\begin{enumerate}

\item \textit{Granular assessment:} Traditional matrices such as the citations count, $h$-index, etc. present an overall impact on a researcher and do not have the capability to capture the ongoing progress and advancement of their work,  whereas $K_z$  can acquire a more nuanced and thorough comprehension of a researcher's contributions as they evolve over time.

\item \textit{Differentiation among researchers:} Even if two researchers possess the same $h$-index, their patterns of impact over time may vary significantly. Analyzing their continuous research impact can uncover disparities in productivity and can provide a more comprehensive understanding of their individual profiles. Hence, $K_z$ allows for a more nuanced differentiation among researchers.

\item \textit{Evaluation of long-term impact:} Researchers may experience fluctuations in their productivity and impact over their careers. Measuring continuous research impact enables the evaluation of long-term contributions. $K_z$ has the capability of highlighting researchers who consistently generate influential work and have a lasting impact on their field.

\item \textit{Career progression and funding decisions:} Many academic institutions, funding agencies, and hiring committees rely on research performance metrics to make decisions. $K_z$ can provide more informed evaluations of researchers, enabling fairer assessments and enhancing the recognition of sustained excellence.

\item \textit{Identification of emerging talent:} Continuous research impact measurement can help identify early-career researchers with promising trajectories. By recognizing their continuous growth and impact, further opportunities can be provided to nurture their potential.
\end{enumerate}

\section{Case studies of $K_z$}

We conducted four case studies to explore the significance of $K_z$. Each case study involved two researchers, namely $R1$ and $R2$. The number of publications was kept constant across all cases, while the focus was on comparing the $h$-index and total citations ($TC$) of two researchers.

\begin{enumerate}
\item \textbf{Case I - Identical $h$-index and total citations:} 
Table~\ref{Table: Case1} represents the first case study where we assumed that both researchers $R1$ and $R2$ have the same $h$-index and total citations count. However, despite sharing these characteristics, researcher $R2$ obtained a higher $K_z$ score than $R1$. This difference in $Kz$ scores can be attributed to the impact of the publication year, which played a dominant role in determining the continuous research impact of each researcher. It highlights the significance of considering the temporal aspect of research contributions when assessing the research impact on individuals.

\item \textbf{Case II - Identical $h$-index and different total citations:} 
In this case (Table~\ref{Table: Case2}), both researchers $R1$ and $R2$ have an equal number of publications and $h$-index, but they differ in their total citations count. Researcher $R1$ has one highly cited paper, while researcher $R2$ has multiple highly cited papers. Despite $R1$ having a higher total number of citations compared to $R2$, $R2$ obtains a higher $K_z$ score. This indicates that the impact of having multiple highly cited papers outweighs the effect of a single highly cited paper in determining the continuous research impact.

\item \textbf{Case III(a) - Different $h$-index and total citations:} 
In this case (Table~\ref{Table: Case3}), both researchers have an equal number of publications but differ in their $h$-index, number of high impact papers, and total citations. Researcher $R1$ has a higher $h$-index but lower total citation count compared to $R2$. However, despite $R1$ having a lower total citation count, they obtain the highest $K_z$ score. This highlights the importance of considering the continuous research impact captured by $K_z$, which takes into account not only the number of citations but also the publication age and impact of publications.

\item \textbf{Case III(b) - Different $h$-index and Total Citations::} 
In this case (Table~\ref{Table: Case4}), we again considered two researchers with an equal number of publications but different $h$-index, high impact papers, and total citations. Researchers $R1$ had a higher $h$-index and total citation count compared to researcher $R2$. Surprisingly, despite these differences, it was researcher $R2$ who obtained the highest $K_z$ score. This finding suggests that the $K_z$ score takes into account factors beyond just $h$-index and total citations, emphasizing the importance of considering the continuous impact and temporal aspects of research contributions.

\end{enumerate}

\begin{table}[!h]
\centering
\caption{Two researchers with identical $h$-index and total citations.}
\begin{tabular}{|c|lllll|lllll|}
\hline
\textbf{Case I} &
  \multicolumn{5}{c|}{\textbf{Researcher 1, $h = 4$}} &
  \multicolumn{5}{c|}{\textbf{Researcher 2, $h = 4$}} \\ \hline
{S. No} &
  \multicolumn{1}{l|}{\textbf{$P_y$}} &
  \multicolumn{1}{l|}{\textbf{$C$}} &
  \multicolumn{1}{l|}{\textbf{$k$}} &
  \multicolumn{1}{l|}{\textbf{$\Delta t$}} &
  \textbf{$k'$} &
  \multicolumn{1}{l|}{\textbf{$P_y$}} &
  \multicolumn{1}{l|}{\textbf{$C$}} &
  \multicolumn{1}{l|}{\textbf{$k$}} &
  \multicolumn{1}{l|}{\textbf{$\Delta t$}} &
  \textbf{$k'$} \\ \hline
1 &
  \multicolumn{1}{l|}{2014} &
  \multicolumn{1}{l|}{40} &
  \multicolumn{1}{l|}{2.2921} &
  \multicolumn{1}{l|}{9} &
  0.255 &
  \multicolumn{1}{l|}{2014} &
  \multicolumn{1}{l|}{2} &
  \multicolumn{1}{l|}{0.4307} &
  \multicolumn{1}{l|}{9} &
  0.048 \\ \hline
2 &
  \multicolumn{1}{l|}{2015} &
  \multicolumn{1}{l|}{30} &
  \multicolumn{1}{l|}{2.1133} &
  \multicolumn{1}{l|}{8} &
  0.264 &
  \multicolumn{1}{l|}{2015} &
  \multicolumn{1}{l|}{3} &
  \multicolumn{1}{l|}{0.6827} &
  \multicolumn{1}{l|}{8} &
  0.085 \\ \hline
3 &
  \multicolumn{1}{l|}{2016} &
  \multicolumn{1}{l|}{0} &
  \multicolumn{1}{l|}{0} &
  \multicolumn{1}{l|}{7} &
  0 &
  \multicolumn{1}{l|}{2016} &
  \multicolumn{1}{l|}{3} &
  \multicolumn{1}{l|}{0.6827} &
  \multicolumn{1}{l|}{7} &
  0.098 \\ \hline
4 &
  \multicolumn{1}{l|}{2017} &
  \multicolumn{1}{l|}{3} &
  \multicolumn{1}{l|}{0.6827} &
  \multicolumn{1}{l|}{6} &
  0.114 &
  \multicolumn{1}{l|}{2016} &
  \multicolumn{1}{l|}{40} &
  \multicolumn{1}{l|}{2.2921} &
  \multicolumn{1}{l|}{7} &
  0.327 \\ \hline
5 &
  \multicolumn{1}{l|}{2018} &
  \multicolumn{1}{l|}{24} &
  \multicolumn{1}{l|}{1.9747} &
  \multicolumn{1}{l|}{5} &
  0.395 &
  \multicolumn{1}{l|}{2017} &
  \multicolumn{1}{l|}{1} &
  \multicolumn{1}{l|}{0} &
  \multicolumn{1}{l|}{6} &
  0 \\ \hline
6 &
  \multicolumn{1}{l|}{2019} &
  \multicolumn{1}{l|}{1} &
  \multicolumn{1}{l|}{0} &
  \multicolumn{1}{l|}{4} &
  0 &
  \multicolumn{1}{l|}{2018} &
  \multicolumn{1}{l|}{30} &
  \multicolumn{1}{l|}{2.1133} &
  \multicolumn{1}{l|}{5} &
  0.423 \\ \hline
7 &
  \multicolumn{1}{l|}{2020} &
  \multicolumn{1}{l|}{1} &
  \multicolumn{1}{l|}{0} &
  \multicolumn{1}{l|}{3} &
  0 &
  \multicolumn{1}{l|}{2019} &
  \multicolumn{1}{l|}{22} &
  \multicolumn{1}{l|}{1.9206} &
  \multicolumn{1}{l|}{4} &
  0.48 \\ \hline
8 &
  \multicolumn{1}{l|}{2021} &
  \multicolumn{1}{l|}{1} &
  \multicolumn{1}{l|}{0} &
  \multicolumn{1}{l|}{2} &
  0 &
  \multicolumn{1}{l|}{2020} &
  \multicolumn{1}{l|}{0} &
  \multicolumn{1}{l|}{0} &
  \multicolumn{1}{l|}{3} &
  0 \\ \hline
9 &
  \multicolumn{1}{l|}{2022} &
  \multicolumn{1}{l|}{0} &
  \multicolumn{1}{l|}{0} &
  \multicolumn{1}{l|}{1} &
  0 &
  \multicolumn{1}{l|}{2021} &
  \multicolumn{1}{l|}{1} &
  \multicolumn{1}{l|}{0} &
  \multicolumn{1}{l|}{2} &
  0 \\ \hline
10 &
  \multicolumn{1}{l|}{2022} &
  \multicolumn{1}{l|}{10} &
  \multicolumn{1}{l|}{1.4307} &
  \multicolumn{1}{l|}{1} &
  1.431 &
  \multicolumn{1}{l|}{2022} &
  \multicolumn{1}{l|}{8} &
  \multicolumn{1}{l|}{1.2921} &
  \multicolumn{1}{l|}{1} &
  1.292 \\ \hline
 &
  \multicolumn{5}{c|}{$TC = 110$, $Kz = 2.459$} &
  \multicolumn{5}{c|}{$TC = 110$, $Kz = 2.753$} \\ \hline
\end{tabular}
\label{Table: Case1}
\end{table}


\begin{table}[!h]
\centering
\caption{Two researchers with identical $h$-index and different total citations.}

\begin{tabular}{|c|lllll|lllll|}
\hline
\textbf{Case II} &
  \multicolumn{5}{c|}{\textbf{Researcher 1, $h=4$}} &
  \multicolumn{5}{c|}{\textbf{Researcher 2, $h=4$}} \\ \hline
{S. No} &
  \multicolumn{1}{l|}{\textbf{$P_y$}} &
  \multicolumn{1}{l|}{\textbf{$C$}} &
  \multicolumn{1}{l|}{\textbf{$k$}} &
  \multicolumn{1}{l|}{\textbf{$\Delta t$}} &
  \textbf{$k'$}&
   \multicolumn{1}{l|}{\textbf{$P_y$}} &
  \multicolumn{1}{l|}{\textbf{$C$}} &
  \multicolumn{1}{l|}{\textbf{$k$}} &
  \multicolumn{1}{l|}{\textbf{$\Delta t$}} &
  \textbf{$k'$} \\ \hline
1 &
  \multicolumn{1}{l|}{2014} &
  \multicolumn{1}{l|}{1000} &
  \multicolumn{1}{l|}{4.2921} &
  \multicolumn{1}{l|}{9} &
  0.477 &
  \multicolumn{1}{l|}{2014} &
  \multicolumn{1}{l|}{500} &
  \multicolumn{1}{l|}{3.8614} &
  \multicolumn{1}{l|}{9} &
  0.429 \\ \hline
2 &
  \multicolumn{1}{l|}{2015} &
  \multicolumn{1}{l|}{4} &
  \multicolumn{1}{l|}{0.8614} &
  \multicolumn{1}{l|}{8} &
  0.108 &
  \multicolumn{1}{l|}{2015} &
  \multicolumn{1}{l|}{300} &
  \multicolumn{1}{l|}{3.544} &
  \multicolumn{1}{l|}{8} &
  0.443 \\ \hline
3 &
  \multicolumn{1}{l|}{2016} &
  \multicolumn{1}{l|}{0} &
  \multicolumn{1}{l|}{0} &
  \multicolumn{1}{l|}{7} &
  0 &
  \multicolumn{1}{l|}{2016} &
  \multicolumn{1}{l|}{100} &
  \multicolumn{1}{l|}{2.8614} &
  \multicolumn{1}{l|}{7} &
  0.409 \\ \hline
4 &
  \multicolumn{1}{l|}{2017} &
  \multicolumn{1}{l|}{4} &
  \multicolumn{1}{l|}{0.8614} &
  \multicolumn{1}{l|}{6} &
  0.144 &
  \multicolumn{1}{l|}{2016} &
  \multicolumn{1}{l|}{0} &
  \multicolumn{1}{l|}{0} &
  \multicolumn{1}{l|}{7} &
  0 \\ \hline
5 &
  \multicolumn{1}{l|}{2018} &
  \multicolumn{1}{l|}{5} &
  \multicolumn{1}{l|}{1.0001} &
  \multicolumn{1}{l|}{5} &
  0.2 &
  \multicolumn{1}{l|}{2017} &
  \multicolumn{1}{l|}{2} &
  \multicolumn{1}{l|}{0.4307} &
  \multicolumn{1}{l|}{6} &
  0.072 \\ \hline
6 &
  \multicolumn{1}{l|}{2019} &
  \multicolumn{1}{l|}{1} &
  \multicolumn{1}{l|}{0} &
  \multicolumn{1}{l|}{4} &
  0 &
  \multicolumn{1}{l|}{2018} &
  \multicolumn{1}{l|}{50} &
  \multicolumn{1}{l|}{2.4307} &
  \multicolumn{1}{l|}{5} &
  0.486 \\ \hline
7 &
  \multicolumn{1}{l|}{2020} &
  \multicolumn{1}{l|}{1} &
  \multicolumn{1}{l|}{0} &
  \multicolumn{1}{l|}{3} &
  0 &
  \multicolumn{1}{l|}{2019} &
  \multicolumn{1}{l|}{1} &
  \multicolumn{1}{l|}{0} &
  \multicolumn{1}{l|}{4} &
  0 \\ \hline
8 &
  \multicolumn{1}{l|}{2021} &
  \multicolumn{1}{l|}{1} &
  \multicolumn{1}{l|}{0} &
  \multicolumn{1}{l|}{2} &
  0 &
  \multicolumn{1}{l|}{2020} &
  \multicolumn{1}{l|}{3} &
  \multicolumn{1}{l|}{0.6827} &
  \multicolumn{1}{l|}{3} &
  0.228 \\ \hline
9 &
  \multicolumn{1}{l|}{2022} &
  \multicolumn{1}{l|}{0} &
  \multicolumn{1}{l|}{0} &
  \multicolumn{1}{l|}{1} &
  0 &
  \multicolumn{1}{l|}{2021} &
  \multicolumn{1}{l|}{1} &
  \multicolumn{1}{l|}{0} &
  \multicolumn{1}{l|}{2} &
  0 \\ \hline
10 &
  \multicolumn{1}{l|}{2022} &
  \multicolumn{1}{l|}{0} &
  \multicolumn{1}{l|}{0} &
  \multicolumn{1}{l|}{1} &
  0 &
  \multicolumn{1}{l|}{2022} &
  \multicolumn{1}{l|}{0} &
  \multicolumn{1}{l|}{0} &
  \multicolumn{1}{l|}{1} &
  0 \\ \hline
 &
  \multicolumn{5}{c|}{$TC = 1016$, $K_z   = 0.929$} &
  \multicolumn{5}{c|}{$TC = 957$, $K_z =   2.067$} \\ \hline
\end{tabular}
\label{Table: Case2}
\end{table}

\begin{table}[!h]
\centering
\caption{Two researchers with different $h$-index and total citations where R1 has higher $h$-index and lower total citations than R2.}
\begin{tabular}{|c|lllll|lllll|}
\hline
\textbf{Case III} &
  \multicolumn{5}{c|}{\textbf{Researcher 1, $h=5$}} &
  \multicolumn{5}{c|}{\textbf{Researcher   2, $h=3$}} \\ \hline
S. No &
 \multicolumn{1}{l|}{\textbf{$P_y$}} &
  \multicolumn{1}{l|}{\textbf{$C$}} &
  \multicolumn{1}{l|}{\textbf{$k$}} &
  \multicolumn{1}{l|}{\textbf{$\Delta t$}} &
  \textbf{$k'$} &
 \multicolumn{1}{l|}{\textbf{$P_y$}} &
  \multicolumn{1}{l|}{\textbf{$C$}} &
  \multicolumn{1}{l|}{\textbf{$k$}} &
  \multicolumn{1}{l|}{\textbf{$\Delta t$}} &
  \textbf{$k'$} \\ \hline
1 &
  \multicolumn{1}{l|}{2014} &
  \multicolumn{1}{l|}{90} &
  \multicolumn{1}{l|}{2.5114} &
  \multicolumn{1}{l|}{9} &
  0.279 &
  \multicolumn{1}{l|}{2014} &
  \multicolumn{1}{l|}{250} &
  \multicolumn{1}{l|}{3.9829} &
  \multicolumn{1}{l|}{9} &
  0.443 \\ \hline
2 &
  \multicolumn{1}{l|}{2015} &
  \multicolumn{1}{l|}{80} &
  \multicolumn{1}{l|}{2.4457} &
  \multicolumn{1}{l|}{8} &
  0.306 &
  \multicolumn{1}{l|}{2015} &
  \multicolumn{1}{l|}{2} &
  \multicolumn{1}{l|}{0.5001} &
  \multicolumn{1}{l|}{8} &
  0.063 \\ \hline
3 &
  \multicolumn{1}{l|}{2016} &
  \multicolumn{1}{l|}{20} &
  \multicolumn{1}{l|}{1.672} &
  \multicolumn{1}{l|}{7} &
  0.239 &
  \multicolumn{1}{l|}{2016} &
  \multicolumn{1}{l|}{2} &
  \multicolumn{1}{l|}{0.5001} &
  \multicolumn{1}{l|}{7} &
  0.071 \\ \hline
4 &
  \multicolumn{1}{l|}{2017} &
  \multicolumn{1}{l|}{3} &
  \multicolumn{1}{l|}{0.6132} &
  \multicolumn{1}{l|}{6} &
  0.102 &
  \multicolumn{1}{l|}{2016} &
  \multicolumn{1}{l|}{82} &
  \multicolumn{1}{l|}{3.1788} &
  \multicolumn{1}{l|}{7} &
  0.454 \\ \hline
5 &
  \multicolumn{1}{l|}{2018} &
  \multicolumn{1}{l|}{24} &
  \multicolumn{1}{l|}{1.7738} &
  \multicolumn{1}{l|}{5} &
  0.355 &
  \multicolumn{1}{l|}{2017} &
  \multicolumn{1}{l|}{2} &
  \multicolumn{1}{l|}{0.5001} &
  \multicolumn{1}{l|}{6} &
  0.083 \\ \hline
6 &
  \multicolumn{1}{l|}{2019} &
  \multicolumn{1}{l|}{2} &
  \multicolumn{1}{l|}{0.3869} &
  \multicolumn{1}{l|}{4} &
  0.097 &
  \multicolumn{1}{l|}{2018} &
  \multicolumn{1}{l|}{110} &
  \multicolumn{1}{l|}{3.3907} &
  \multicolumn{1}{l|}{5} &
  0.678 \\ \hline
7 &
  \multicolumn{1}{l|}{2020} &
  \multicolumn{1}{l|}{3} &
  \multicolumn{1}{l|}{0.6132} &
  \multicolumn{1}{l|}{3} &
  0.204 &
  \multicolumn{1}{l|}{2019} &
  \multicolumn{1}{l|}{1} &
  \multicolumn{1}{l|}{0} &
  \multicolumn{1}{l|}{4} &
  0 \\ \hline
8 &
  \multicolumn{1}{l|}{2021} &
  \multicolumn{1}{l|}{3} &
  \multicolumn{1}{l|}{0.6132} &
  \multicolumn{1}{l|}{2} &
  0.307 &
  \multicolumn{1}{l|}{2020} &
  \multicolumn{1}{l|}{2} &
  \multicolumn{1}{l|}{0.5001} &
  \multicolumn{1}{l|}{3} &
  0.167 \\ \hline
9 &
  \multicolumn{1}{l|}{2022} &
  \multicolumn{1}{l|}{2} &
  \multicolumn{1}{l|}{0.3869} &
  \multicolumn{1}{l|}{1} &
  0.387 &
  \multicolumn{1}{l|}{2021} &
  \multicolumn{1}{l|}{2} &
  \multicolumn{1}{l|}{0.5001} &
  \multicolumn{1}{l|}{2} &
  0.25 \\ \hline
10 &
  \multicolumn{1}{l|}{2022} &
  \multicolumn{1}{l|}{23} &
  \multicolumn{1}{l|}{1.75} &
  \multicolumn{1}{l|}{1} &
  1.75 &
  \multicolumn{1}{l|}{2022} &
  \multicolumn{1}{l|}{0} &
  \multicolumn{1}{l|}{0} &
  \multicolumn{1}{l|}{1} &
  0 \\ \hline
 &
  \multicolumn{5}{c|}{$TC = 250$, $K_z = 4.026$} &
  \multicolumn{5}{c|}{$TC = 453$,$ K_z = 2.209$} \\ \hline
\end{tabular}
\label{Table: Case3}
\end{table}

\begin{table}[!h]
\centering
\caption{Two researchers with different $h$-index and total citations where R1 has higher $h$-index and total citations than R2.}
\begin{tabular}{|c|lllll|lllll|}
\hline
\textbf{Case IV} &
  \multicolumn{5}{c|}{\textbf{Researcher 1, $h=6$}} &
  \multicolumn{5}{c|}{\textbf{Researcher 2, $h=4$}} \\ \hline
S. No &
  \multicolumn{1}{l|}{\textbf{$P_y$}} &
  \multicolumn{1}{l|}{\textbf{$C$}} &
  \multicolumn{1}{l|}{\textbf{$k$}} &
  \multicolumn{1}{l|}{\textbf{$\Delta t$}} &
  \textbf{$k'$}  &
 \multicolumn{1}{l|}{\textbf{$P_y$}} &
  \multicolumn{1}{l|}{\textbf{$C$}} &
  \multicolumn{1}{l|}{\textbf{$k$}} &
  \multicolumn{1}{l|}{\textbf{$\Delta t$}} &
  \textbf{$k'$}  \\ \hline
1 &
  \multicolumn{1}{l|}{2014} &
  \multicolumn{1}{l|}{200} &
  \multicolumn{1}{l|}{2.7228} &
  \multicolumn{1}{l|}{9} &
  0.303 &
  \multicolumn{1}{l|}{2014} &
  \multicolumn{1}{l|}{2} &
  \multicolumn{1}{l|}{0.4307} &
  \multicolumn{1}{l|}{9} &
  0.048 \\ \hline
2 &
  \multicolumn{1}{l|}{2015} &
  \multicolumn{1}{l|}{150} &
  \multicolumn{1}{l|}{2.575} &
  \multicolumn{1}{l|}{8} &
  0.322 &
  \multicolumn{1}{l|}{2015} &
  \multicolumn{1}{l|}{2} &
  \multicolumn{1}{l|}{0.4307} &
  \multicolumn{1}{l|}{8} &
  0.054 \\ \hline
3 &
  \multicolumn{1}{l|}{2016} &
  \multicolumn{1}{l|}{5} &
  \multicolumn{1}{l|}{0.8271} &
  \multicolumn{1}{l|}{7} &
  0.118 &
  \multicolumn{1}{l|}{2016} &
  \multicolumn{1}{l|}{3} &
  \multicolumn{1}{l|}{0.6827} &
  \multicolumn{1}{l|}{7} &
  0.098 \\ \hline
4 &
  \multicolumn{1}{l|}{2017} &
  \multicolumn{1}{l|}{10} &
  \multicolumn{1}{l|}{1.1833} &
  \multicolumn{1}{l|}{6} &
  0.197 &
  \multicolumn{1}{l|}{2016} &
  \multicolumn{1}{l|}{1} &
  \multicolumn{1}{l|}{0} &
  \multicolumn{1}{l|}{7} &
  0 \\ \hline
5 &
  \multicolumn{1}{l|}{2018} &
  \multicolumn{1}{l|}{35} &
  \multicolumn{1}{l|}{1.8271} &
  \multicolumn{1}{l|}{5} &
  0.365 &
  \multicolumn{1}{l|}{2017} &
  \multicolumn{1}{l|}{280} &
  \multicolumn{1}{l|}{3.5011} &
  \multicolumn{1}{l|}{6} &
  0.584 \\ \hline
6 &
  \multicolumn{1}{l|}{2019} &
  \multicolumn{1}{l|}{1} &
  \multicolumn{1}{l|}{0} &
  \multicolumn{1}{l|}{4} &
  0 &
  \multicolumn{1}{l|}{2018} &
  \multicolumn{1}{l|}{2} &
  \multicolumn{1}{l|}{0.4307} &
  \multicolumn{1}{l|}{5} &
  0.086 \\ \hline
7 &
  \multicolumn{1}{l|}{2020} &
  \multicolumn{1}{l|}{33} &
  \multicolumn{1}{l|}{1.7969} &
  \multicolumn{1}{l|}{3} &
  0.599 &
  \multicolumn{1}{l|}{2019} &
  \multicolumn{1}{l|}{40} &
  \multicolumn{1}{l|}{2.2921} &
  \multicolumn{1}{l|}{4} &
  0.573 \\ \hline
8 &
  \multicolumn{1}{l|}{2021} &
  \multicolumn{1}{l|}{1} &
  \multicolumn{1}{l|}{0} &
  \multicolumn{1}{l|}{2} &
  0 &
  \multicolumn{1}{l|}{2020} &
  \multicolumn{1}{l|}{70} &
  \multicolumn{1}{l|}{2.6398} &
  \multicolumn{1}{l|}{3} &
  0.88 \\ \hline
9 &
  \multicolumn{1}{l|}{2022} &
  \multicolumn{1}{l|}{2} &
  \multicolumn{1}{l|}{0.3563} &
  \multicolumn{1}{l|}{1} &
  0.356 &
  \multicolumn{1}{l|}{2021} &
  \multicolumn{1}{l|}{2} &
  \multicolumn{1}{l|}{0.4307} &
  \multicolumn{1}{l|}{2} &
  0.215 \\ \hline
10 &
  \multicolumn{1}{l|}{2022} &
  \multicolumn{1}{l|}{32} &
  \multicolumn{1}{l|}{1.7811} &
  \multicolumn{1}{l|}{1} &
  1.781 &
  \multicolumn{1}{l|}{2022} &
  \multicolumn{1}{l|}{50} &
  \multicolumn{1}{l|}{2.4307} &
  \multicolumn{1}{l|}{1} &
  2.431 \\ \hline
 &
  \multicolumn{5}{c|}{$TC = 469$, $K_z = 4.041$} &
  \multicolumn{5}{c|}{$TC = 452$, $K_z = 4.969$} \\ \hline
\end{tabular}
\label{Table: Case4}
\end{table}

\section{Empirical study}

To calculate the continuous research impact ($K_z$) of researchers, the research profiles of 376 individuals affiliated with Monash University, Australia were obtained. Monash University is a public research institution located in Australia, and information about the researchers can be found on their webpage at \url{https://research.monash.edu/en/persons/}. The webpage provides the researcher's research ID and Orcid ID, which facilitated the extraction of their publication details and citations from the Web of Science database. From a pool of 6316 researchers' profiles, we selected 376 profiles across different disciplines, ensuring a range of $h$-index values ($1 \leq h \leq 112$). The choice of databases was made based on data availability. For each researcher ID, information regarding the publication year and the corresponding citations received were extracted. For each researcher, the $h$-index, $g$-index, and $K_z$ were computed. Additionally, the overall research age or career length of the researcher was determined by subtracting the year of his/her first publication from the current year.

\subsection{Comparison of $K_z$ with $h$-index and career length}
\begin{figure}[!h]
    \centering
\includegraphics[width=0.65\linewidth]{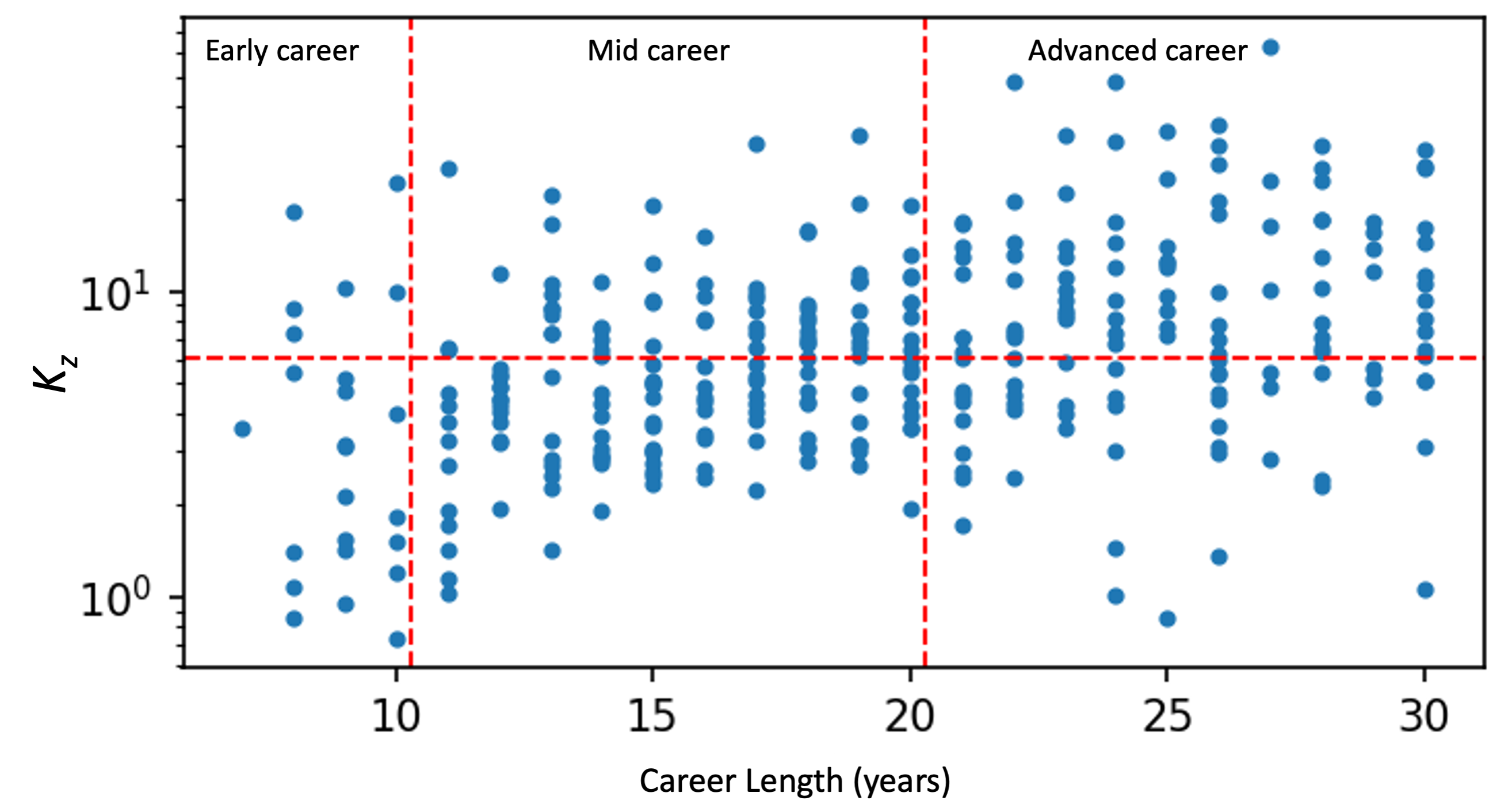} 
    \caption{Scattered plot of $K_z$ versus career length. Each dot correspond to a researcher. The horizontal dashed line represents the median of the axis and vertical dashed lines divides the plots in three zones based on the researcher's career length.}
    \label{fig:2}
\end{figure}

By using equation~\ref{eq3}, we calculated the $K_z$ score of 376 researchers. In Figure~\ref{fig:2}, a scatter plot depicting the relationship between $K_z$ and career length. Each dot on the plot represents an individual researcher. The horizontal dashed line represents the median of the axis, while vertical dashed lines are used to divide the plot into three zones based on the length of the researchers' careers: early career ($\leq 10$ years), mid career ($>10$ and $\leq 20$ years) and advanced career ($>20$ years). This visualization clearly differentiate between the star performer and average performer at different career stages. 

Table~\ref{Table: Same-h} provides examples of researchers who have the same $h$-index values of 25 and 30. It also includes the computation of the $g$-index, which demonstrates that researchers with the same $h$-index can have different $g$ values due to variations in their total citation counts. Therefore, it is possible for a researcher with a lower citation count to have a higher $g$-index, and vice versa. The presence of the same $h$-index highlights its limitation in differentiating the top-performing researcher from others whereas $K_z$ significantly differentiates the impactful researcher from others.  This distinction highlights the varying impact among researchers. Similarly, Table~\ref{Table: Same-age} showcases profiles of researchers with the same career age, yet their $K_z$ scores differ. $K_z$ clearly differentiates the impactful researcher from others where researchers are of the same career length. The same observation applies to total publication and citation counts.

In Table~\ref{Table:Comparison}, we examined 11 comparative cases of researchers with identical $h$-index and career length. Among these cases, one noteworthy instance is $S1$, where two researchers share the same career length of 8 years and $h$-index of 12. However, the researcher with higher total publications and citations count has a higher $K_z$ score than the other.  Whereas, in case $S3$, two researchers have a career length of 13 years and $h$-index of 19, the one with lower total publications but higher citation counts, compared to the other researcher, has a higher $K_z$ score. On the other hand, in case $S7$, two researchers have a career length of 17 years and $h$-index of 13, the one with higher total publications but lower citation counts, compared to the other researcher, has a higher $K_z$ score. Hence, this indicates that the $K_z$ metric considers all relevant research indicators such as total publications, citation count, $h$-index, and publication age to capture the continuous impact of an individual. It is not safe to assume that a higher $K_z$ score is solely determined by either higher total publications or higher citation counts. Additionally, it cannot be concluded that a person with a higher $h$-index will always have a higher $K_z$ score. The $K_z$ metric takes a comprehensive approach in evaluating research impact, considering multiple factors simultaneously.


\begin{table}[h!]
\caption{Comparative analysis among researchers having identical $h$-index.}
\begin{tabular}{|c|c|c|c|c|c|c|}
\hline
\multicolumn{1}{|l|}{\textbf{WoS Researcher ID}} &
  \textbf{\begin{tabular}[c]{@{}c@{}}Career \\ Length (yrs)\end{tabular}} &
  \textbf{Publications} &
  \textbf{\begin{tabular}[c]{@{}c@{}}Total \\ Citations\end{tabular}} &
  \textbf{$h$-index} &
  \textbf{$g$-index} &
  \textbf{$K_z$} \\ \hline
B-6419-2008 & 17 & 44  & 2415 & 25 & 44 & 5.24   \\ \hline
H-6054-2014 & 19 & 38  & 3433 & 25 & 37 & 6.76   \\ \hline
D-5776-2019 & 26 & 68  & 1984 & 25 & 45 & 6.828  \\ \hline
J-1532-2014 & 18 & 59  & 2982 & 25 & 46 & 7.896  \\ \hline
N-8153-2014 & 20 & 78  & 4217 & 25 & 65 & 9.156  \\ \hline
E-6623-2015 & 14 & 59  & 1530 & 25 & 38 & 10.618 \\ \hline
A-3854-2010 & 21 & 86  & 2034 & 25 & 44 & 11.224 \\ \hline
K-5277-2012 & 24 & 73  & 3783 & 25 & 64 & 11.912 \\ \hline
B-8486-2008 & 29 & 79  & 2851 & 30 & 54 & 4.487  \\ \hline
G-1412-2012 & 34 & 69  & 2816 & 30 & 56 & 5.517  \\ \hline
H-3196-2013 & 13 & 94  & 2538 & 30 & 49 & 8.684  \\ \hline
F-2273-2010 & 16 & 102 & 2627 & 30 & 48 & 10.446 \\ \hline
I-1956-2014 & 23 & 123 & 3797 & 30 & 60 & 11.05  \\ \hline
I-1738-2013 & 19 & 105 & 3306 & 30 & 57 & 11.309 \\ \hline
D-4239-2011 & 25 & 133 & 3343 & 30 & 59 & 12.475 \\ \hline
H-4935-2013 & 15 & 100 & 2945 & 30 & 52 & 18.97  \\ \hline
\end{tabular}
\label{Table: Same-h}
\end{table}

\begin{table}[h!]
\caption{Comparative analysis among researchers having identical career length.}
\begin{tabular}{|c|c|c|c|c|c|c|}
\hline
\multicolumn{1}{|l|}{\textbf{WoS Researcher ID}} &
  \textbf{\begin{tabular}[c]{@{}c@{}}Career \\ Length (yrs)\end{tabular}} &
  \textbf{Publications} &
  \textbf{\begin{tabular}[c]{@{}c@{}}Total \\ Citations\end{tabular}} &
  \textbf{$h$-index} &
  \textbf{$g$-index} &
  \textbf{$K_z$} \\ \hline

K-5514-2018 					  & 10 & 9   & 32    & 4  & 6   & 1.043  \\ \hline
P-7354-2019                       & 10 & 8   & 171   & 6  & 8   & 1.69   \\ \hline
I-9365-2017                       & 10 & 20  & 287   & 10 & 17  & 3.823  \\ \hline
G-3877-2013                       & 10 & 75  & 1189  & 18 & 34  & 9.813  \\ \hline
L-4481-2018                       & 10 & 90  & 6012  & 28 & 83  & 22.385 \\ \hline
N-4364-2019                       & 20 & 23  & 757   & 14 & 23  & 1.905  \\ \hline
A-4190-2009                       & 20 & 32  & 832   & 14 & 29  & 3.795  \\ \hline
B-7556-2008                       & 20 & 60  & 7144  & 27 & 54  & 6.847  \\ \hline
C-9764-2013                       & 20 & 122 & 5917  & 42 & 77  & 10.995 \\ \hline
I-1587-2014                       & 20 & 107 & 1127  & 18 & 30  & 12.88  \\ \hline
C-4319-2011                       & 20 & 170 & 5080  & 39 & 62  & 19.088 \\ \hline
H-9193-2014                       & 30 & 26  & 181   & 8  & 14  & 2.939  \\ \hline
P-8366-2016                       & 30 & 98  & 5701  & 40 & 77  & 6.378  \\ \hline
B-9553-2008                       & 30 & 91  & 6784  & 45 & 85  & 10.524 \\ \hline
H-5706-2014                       & 30 & 171 & 4559  & 35 & 60  & 15.996 \\ \hline
A-5452-2008                       & 30 & 283 & 26495 & 89 & 158 & 25.657 \\ \hline
I-6251-2012                       & 30 & 280 & 58171 & 68 & 244 & 29.05  \\ \hline
\end{tabular}
\label{Table: Same-age}
\end{table}

\begin{table}[h!]
\caption{Comparative analysis among researchers having identical research career length (yrs) and $h$-index. }
\begin{tabular}{|c|l|c|c|c|c|c|c|}
\hline
\textbf{S.No} &
\textbf{WoS Researcher ID} &
  \textbf{\begin{tabular}[c]{@{}c@{}}Career \\ Length (yrs)\end{tabular}} &
  \textbf{Publications} &
  \textbf{\begin{tabular}[c]{@{}c@{}}Total \\ Citations\end{tabular}} &
  \textbf{$h$-index} &
  \textbf{$g$-index} &
  \textbf{$K_z$} \\ \hline
\multirow{2}{*}{S1}  & F-9424-2013   & 8  & 37  & 1595 & 12 & 5  & 7.041  \\ \cline{2-8} 
                     & O-7942-2018   & 8  & 34  & 454  & 12 & 7  & 5.291  \\ \hline
\multirow{2}{*}{S2}  & AAE-7279-2019 & 12 & 47  & 1529 & 15 & 18 & 11.122 \\ \cline{2-8} 
                     & I-9929-2012   & 12 & 37  & 1236 & 15 & 26 & 4.321  \\ \hline
\multirow{2}{*}{S3}  & L-4989-2018   & 13 & 84  & 1875 & 19 & 16 & 20.182 \\ \cline{2-8} 
                     & M-7607-2014   & 13 & 106 & 1130 & 19 & 29 & 8.26   \\ \hline
\multirow{2}{*}{S4}  & E-6431-2011   & 14 & 16  & 508  & 8  & 16 & 4.057  \\ \cline{2-8} 
                     & N-1676-2017   & 14 & 14  & 726  & 8  & 22 & 2.771  \\ \hline
\multirow{2}{*}{S5}  & A-7222-2013   & 14 & 28  & 608  & 14 & 21 & 6.299  \\ \cline{2-8} 
                     & L-1320-2019   & 14 & 23  & 875  & 14 & 21 & 3.264  \\ \hline
\multirow{2}{*}{S6}  & K-7419-2014   & 15 & 52  & 482  & 11 & 16 & 2.845  \\ \cline{2-8} 
                     & G-1470-2011   & 15 & 36  & 351  & 11 & 13 & 4.741  \\ \hline
\multirow{2}{*}{S7}  & O-9174-2014   & 17 & 36  & 708  & 13 & 22 & 4.444  \\ \cline{2-8} 
                     & J-5651-2016   & 17 & 16  & 857  & 13 & 22 & 2.173  \\ \hline
\multirow{2}{*}{S8}  & Q-9068-2018   & 18 & 47  & 2034 & 21 & 36 & 7.279  \\ \cline{2-8} 
                     & H-4554-2014   & 18 & 53  & 1462 & 21 & 26 & 8.99   \\ \hline
\multirow{2}{*}{S9}  & F-6776-2014   & 18 & 159 & 1843 & 23 & 28 & 15.62  \\ \cline{2-8} 
                     & H-8387-2012   & 18 & 78  & 1798 & 23 & 34 & 8.635  \\ \hline
\multirow{2}{*}{S10} & F-4112-2014   & 22 & 18  & 617  & 13 & 38 & 2.402  \\ \cline{2-8} 
                     & C-6296-2014   & 22 & 35  & 1456 & 13 & 38 & 4.842  \\ \hline
\multirow{2}{*}{S11} & C-2440-2013   & 28 & 38  & 6087 & 27 & 60 & 2.401  \\ \cline{2-8} 
                     & N-5018-2017   & 28 & 87  & 2588 & 27 & 62 & 7.02   \\ \hline
\end{tabular}

\label{Table:Comparison}
\end{table}

\subsection{Probability distribution of $K_z$}
\begin{figure}[!h]
    \centering
\includegraphics[width=0.75\linewidth]{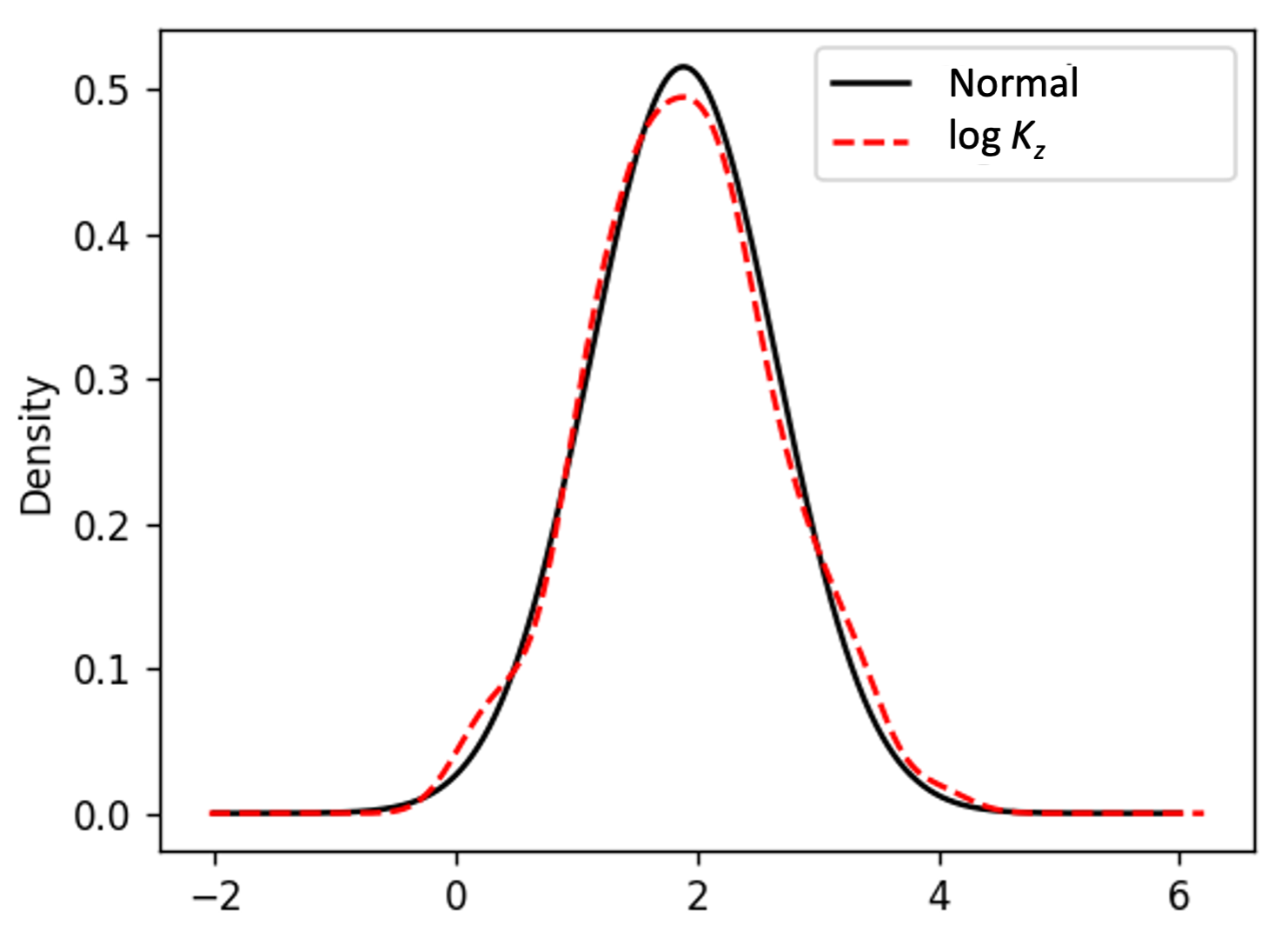} 
    \caption{Distribution of $\log (K_z)$ (dashed red) versus normal distribution (solid black) with same $\mu$ and $\sigma$.}
    \label{fig:1}
\end{figure}

Figure~\ref{fig:1} presents a graphical representation of the plot for $\log(K_z)$, which exhibits a mean value of $\mu$ and a standard deviation of $\sigma$. This plot is compared to the normal distribution with the same mean and standard deviation. The overlapping nature of the two plots suggests that the variable $K_z$ follows a log-normal distribution. To confirm this observation, a ``Goodness of Fit" test was conducted using the $\chi^2$ distribution. The objective of the Goodness of Fit Test was to assess the suitability of the null hypothesis that states ``the distribution of $\log(K_z)$ conforms well to a normal distribution.'' The test was executed in the following manner:

The logarithm of the values of $K_z$ was computed, and these values were then classified into seven distinct classes, taking into account the mean ($\mu = 0.78787$) and standard deviation ($\sigma = 0.37448$). Subsequently, the observed frequencies ($O_i$) for each class were determined. To obtain the expected frequencies ($E_i$), the entire dataset consisting of 376 observations was subjected to calculations based on the normal distribution. The specific calculations and their results are provided in Table~\ref{GoF}.

The $\chi^2$ value was computed using the formula $\chi^2 = \sum \frac{(O_i - E_i)^2}{E_i}$ and yielded a value of $7.466$. As the calculated $\chi^2$ value is smaller than the critical value $\chi^2_{(6,0.05)} = 12.592$, we cannot reject the null hypothesis at a significance level of $0.05$. Therefore, we can conclude that $\log(K_z)$ is a suitable fit for the normal distribution.

\subsubsection{Identification of top contributors and low contributors}

In the case of a normal distribution, the middle 50\% of the data is encompassed within a range of $+0.67$ and $-0.67$ standard scores from the mean. Consequently, researchers in the top 25\% satisfy the condition $K_z \geq e^{(\mu+ 0.67 \sigma)}$, while researchers in the bottom 25\% satisfy the condition $K_z \leq e^{(\mu - 0.67 \sigma)}$. 
Similarly, using the properties of normal distribution, the $\alpha \%$ of top and bottom performers can be identifies.
Unlike previous indices such as the $h$, $g$, $e$, $h_c$, etc., the $K_z$-index allows for the identification of both top and bottom contributors. This categorization based on $K_z$ scores can be beneficial for universities, scientific communities, and research funding agencies in identifying significant contributors.

\begin{table}[h!]
\caption{Goodness of fit test.}
\begin{tabular}{|c|c|c|}
\hline
\textbf{Classes} & \textbf{\begin{tabular}[c]{@{}c@{}}Observed \\ Frequencies $(O_i)$\end{tabular}} & \textbf{\begin{tabular}[c]{@{}c@{}}Expected frequencies \\ $(E_i)$ for$\mathcal{N} (\mu, \sigma)$\end{tabular}} \\ \hline
$\log(K_z) < \mu  - 1.5 \sigma$     & 14  & 25  \\ \hline
$\mu  - 1.5 \sigma \leq \log(K_z) < \mu  - \sigma$     & 34  & 35  \\ \hline
$\mu  -  \sigma \leq \log(K_z) < \mu  - 0.5 \sigma$     & 57  & 56  \\ \hline
$\mu  - 0.5  \sigma \leq \log(K_z) < \mu + 0.5 \sigma$     & 157 & 144 \\ \hline
$\mu  + 0.5  \sigma \leq \log(K_z) < \mu + \sigma$     & 57  & 56  \\ \hline
$\mu  +  \sigma \leq \log(K_z) < \mu +1.5 \sigma$     & 29  & 35  \\ \hline
$ \log(K_z) \geq \mu -1.5 \sigma$     & 28  & 25  \\ \hline
Total & 376 &     \\ \hline
\end{tabular}
\label{GoF}
\end{table}

\section{Discussion and conclusion}

In this study, we have discussed various research indicators, including total publications, citations count,  $h$-index, $g$-index, etc., commonly used to measure the impact of research. While total publications, citation count, and $h$-index are commonly used indicators to assess research impact, they have some limitations when considered individually.

\begin{enumerate}

\item \textit{Total publications}: Relying solely on the number of publications can be misleading, as it does not consider the quality or impact of those publications. Quantity alone does not reflect the significance or influence of a researcher's work.

\item \textit{Citations count}: While citation count is a useful indicator of the influence and visibility of a researcher's work, it can be influenced by factors such as the field of study, publication age, and citation practices within the research community. Additionally, self-citations can artificially inflate citation counts and impact assessments.

\item \textit{$h$-index: }The $h$-index takes into account both the number of publications and their corresponding citations. However, it does not differentiate between highly cited publications and those with fewer citations. A researcher with a few highly influential papers can have the same $h$-index as someone with many moderately cited papers. Additionally $h$-index ignores all the papers which are cited less than the $h$.

\item \textit{Temporal considerations: }Individual metrics may not capture the continuous progress and development of a researcher's work over time. They provide a snapshot of impact at a specific moment and may not reflect the long-term contributions or evolving research trajectory.

\end{enumerate}
To overcome these limitations and capture the dynamic nature of research impact, it is essential to consider multiple indicators and employ comprehensive assessment approaches like the $K_z$ metric, which incorporates various factors to provide a more nuanced understanding of research impact. $K_z$ is filed independent as well as takes into account the temporal aspect of the work. Unlike other research indicators, $K_z$ takes into account not only the total publications and citations count but the age of the publications too. Our results demonstrate how $K_z$ can effectively differentiate between two potential researchers who may have the same $h$-index, citations count, or career length. By incorporating $K_z$ into the evaluation process, we can better assess the research dynamics of an individual and gain insights into their continuous impact over time.

To conclude, $K_z$ holds the potential to serve as a superior measure for capturing the impact of individuals, institutions, or journals. Its comprehensive consideration of various research indicators allows a more nuanced assessment of research impact. Further$K_z$ can be utilized as a ranking method to evaluate and rank researchers within an institution based on their research impact. Similarly, institutions and journals can be compared and ranked according to their research impact. This information can be valuable in decision-making processes, as funding agencies, research award committees and hiring bodies can leverage the power of $K_z$ to rank potential candidates within a specific field. It provides a standardized tool to assess and compare the impact of research entities, facilitating more informed decisions and promoting recognition based on research excellence.

There are some challenges associated with computing the $K_z$ metric too. Some of the potential challenges include:
\begin{enumerate}
\item \textit{Data availability and accuracy:} Obtaining accurate and comprehensive data from various sources can be a challenge. Different databases may have variations in the coverage of publications and citations, potentially leading to incomplete or inconsistent data.

\item \textit{Data quality and reliability:} The accuracy and reliability of the data sources used for computing $K_z$ are crucial as inaccurate or incomplete data can result in misleading or flawed assessments of research impact.

\item \textit{Self-citation manipulation:} The issue of self-citation manipulation, where researchers excessively cite their own work to inflate their impact metrics, can pose a challenge as detecting such manipulations requires careful scrutiny and data filtering techniques.

\end{enumerate}

As discussed, it can be inferred that the $K_z$ index is a comprehensive mathematical function that considers multiple factors to assess the impact of a researcher. These factors include the researcher's total publications, the citation count of each paper, the researcher's $h$-index, and the age of publication. The $K_z$ index recognizes influential papers which often receive citations at a faster rate, indicating a greater impact, and therefore assigns them higher weight in impact evaluation. By considering these aspects, the $K_z$ index tends to yield higher values in cases where a researcher has made significant contributions that have garnered substantial citations.

\section*{Acknowlegement}

We acknowledge the suggestions provided by Dr. Satyam Mukherjee.

\bibliographystyle{model1-num-names}

\bibliography{cas-refs}

\end{document}